\documentclass[a4paper,11pt]{article}
\usepackage[utf8]{inputenc}

\usepackage{jheppub} 

\usepackage[T1]{fontenc} 

\author{Dhimiter D.~Canko, Costas G. Papadopoulos and Nikolaos Syrrakos}

\affiliation{Institute of Nuclear and Particle Physics, NCSR `Demokritos', Agia Paraskevi, 15310, Greece}


\emailAdd{jimcanko@phys.uoa.gr, costas.papadopoulos@cern.ch, syrrakos@inp.demokritos.gr}

\title{Analytic representation of all  planar two-loop five-point Master Integrals with one off-shell leg}

\abstract{
We present analytic expressions in terms of polylogarithmic functions for all three families of planar two-loop five-point Master Integrals with one off-shell leg. The calculation is based on the Simplified Differential Equations approach. The results are relevant to the study of many $2\to 3$ scattering processes  of interest at the LHC, especially for the leading-color $W+2$ jets production.
}

\begin{document}
\unitlength1cm
\maketitle
\allowdisplaybreaks

\section{Introduction } 
\label{sIntro}

During the last decade we have learned that in order to discover new phenomena in Nature,  
from gravitational wave astronomy~\cite{Abbott:2016blz} to high-energy
physics~\cite{Aad:2012tfa,Chatrchyan:2012xdj},
we need not only very sophisticated, state-of-the-art instrumentation, but also very precise theoretical predictions.
The coming LHC Run 3 and the High Luminosity LHC Run scheduled after it, require the most precise description of the scattering processes under investigation, in order to fully exploit the machine’s potential~\cite{Amoroso:2020lgh,Bendavid:2018nar}. In the future, the FCC (Future Circular Collider) project will also further boost the demands in the direction of precision calculations~\cite{Mangano:2017tke}.   

Next-to-next-to-leading order (NNLO) accuracy is needed for the vast majority of QCD dominated scattering processes at the LHC (see~\cite{Heinrich:2017una} and references therein). 
Over the last years, NNLO QCD corrections for most of the $2\to 2$ processes, including two-jet, top-pair and gauge bosons production, have been completed and already used in phenomenological and experimental studies~\cite{Azzi:2019yne}. 
Two-loop amplitude computations require the reduction of the scattering matrix element in terms of basis integrals, usually referred to as Master Integrals (MI). Traditional reduction techniques based on integration-by-part identities~\cite{Chetyrkin:1981qh,Tkachov:1981wb,Laporta:2001dd} (IBP), {\it at the integral level}, are now more and more replaced by {\it integrand-reduction methods}~\cite{Badger:2012dp,Mastrolia:2012wf,Ita:2015tya}, following the one-loop paradigm~\cite{Ossola:2006us}. Results for five-point two-loop amplitudes, relevant for three-jet/photon, $W,Z,H+2$ jets production have been recently presented~\cite{Badger:2017jhb,Abreu:2017hqn,Abreu:2018jgq,Hartanto:2019uvl,Abreu:2019odu}. Moreover, a complete NNLO calculation for the relatively easy case of three-photon production at the LHC, has been recently published~\cite{Chawdhry:2019bji,Kallweit:2020gcp}. Despite the progress in understanding amplitude reduction and real radiation corrections at NNLO, a remarkable contradistinction with the NLO case is that the basis of Master Integrals at two loops is still far from complete\footnote{For interesting alternative approaches see references~\cite{Anastasiou:2020sdt,Anastasiou:2018rib}.}. 

Multi-loop Master Integrals have been studied for many years now. The most appropriate method to obtain analytic expressions and accurate numerical estimates of multi-scale multi-loop Feynman Integrals is the differential equations (DE) approach~\cite{Kotikov:1990kg,Kotikov:1991pm,Bern:1992em,Remiddi:1997ny,Gehrmann:1999as}. With the introduction of the canonical form of the differential equations~\cite{Henn:2013pwa}, a major step towards the understanding of the mathematical structure of Feynman Integrals and subsequently of the scattering amplitudes has been achieved. The complexity of two-loop Feynman Integrals is determined by the number of internal massive propagators and the number of external particles, i.e. the total number of independent "kinematical" scales involved. Feynman Integrals with a relatively small number of scales satisfy canonical differential equations and can be expressed in terms of multiple (or Goncharov) polylogarithms~\cite{Goncharov:1998kja,Remiddi:1999ew,Goncharov:2001iea}, a class of functions that have been well understood by now.   
Moreover, in the last couple of years, new mathematical structures~\cite{Adams:2015gva,Bonciani:2016qxi,Ablinger:2017bjx,Bourjaily:2017bsb,Broedel:2018iwv} (elliptic polylogarithms) have been studied in order to obtain analytic insight of more complicated Feynman Integrals.
With a complete basis of two-loop Master Integrals, it is hoped that an automation of NNLO calculations for arbitrary scattering processes can be achieved in the near future.

\begin{figure}[t!]
\centering
\includegraphics[width=0.20 \linewidth]{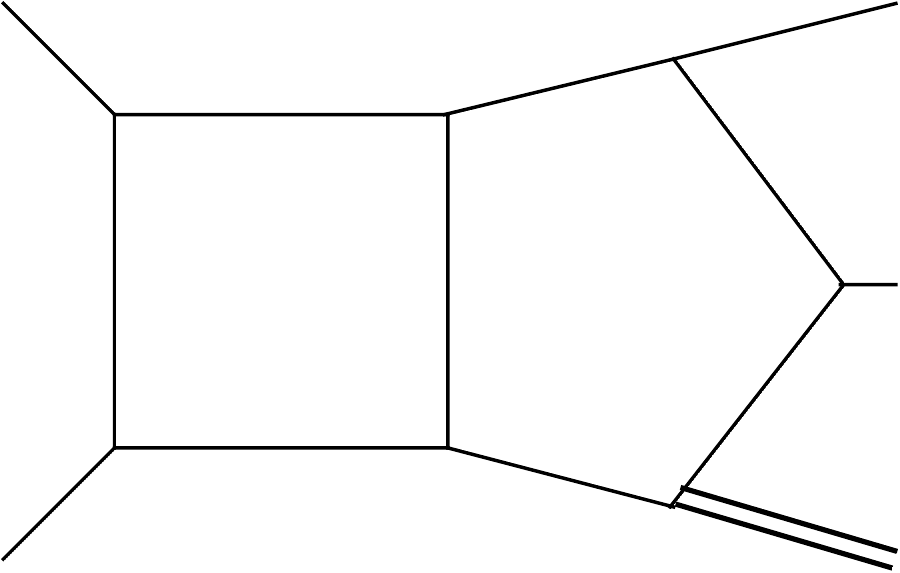} \hspace{0.4 cm}
\includegraphics[width=0.20 \linewidth]{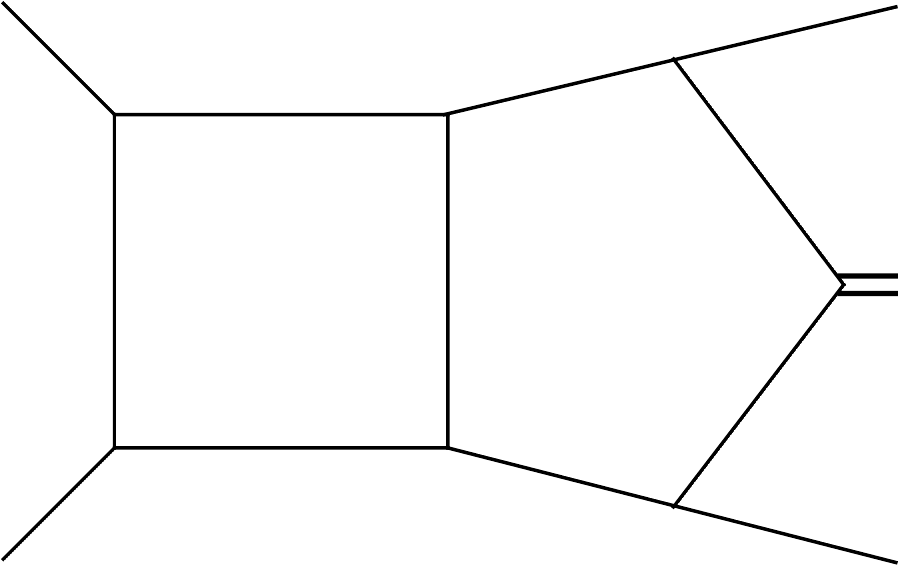} \hspace{0.4 cm}
\includegraphics[width=0.20 \linewidth]{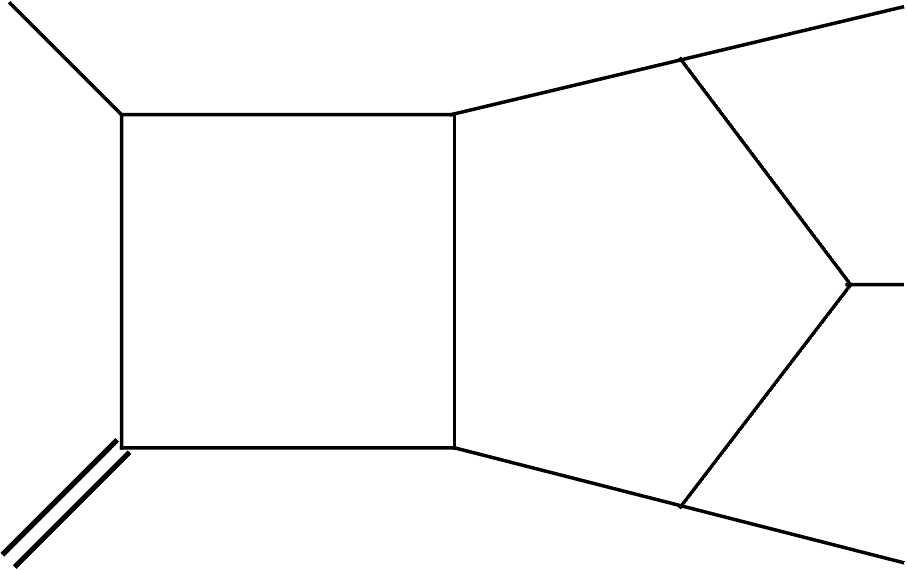}
\\[12pt]
\centering
\includegraphics[width=0.20 \linewidth]{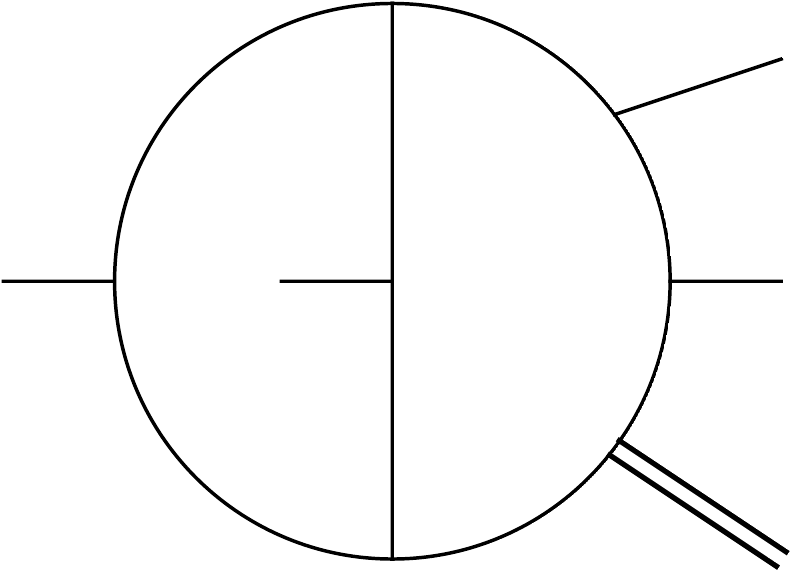} \hspace{0.6 cm}
\includegraphics[width=0.20 \linewidth]{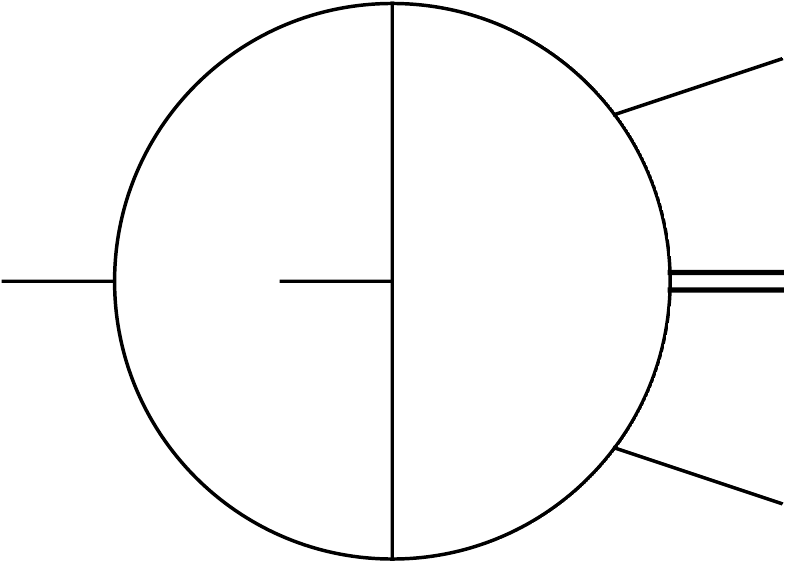} \hspace{0.6 cm}
\includegraphics[width=0.20 \linewidth]{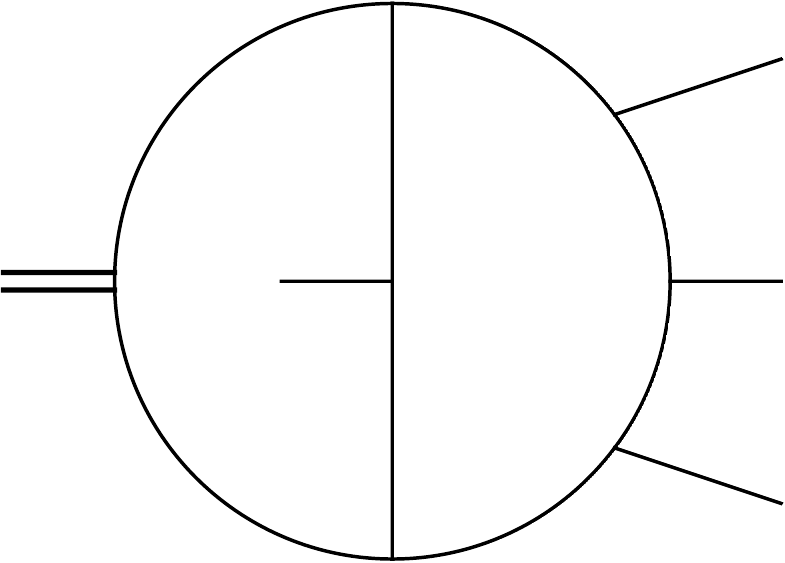} 
\\[12pt]
\includegraphics[width=0.20 \linewidth]{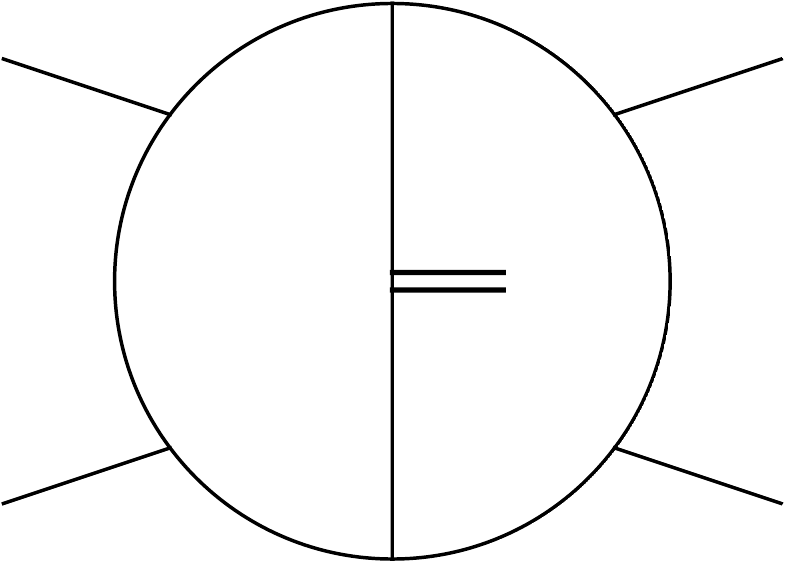} \hspace{0.6 cm}
\includegraphics[width=0.20 \linewidth]{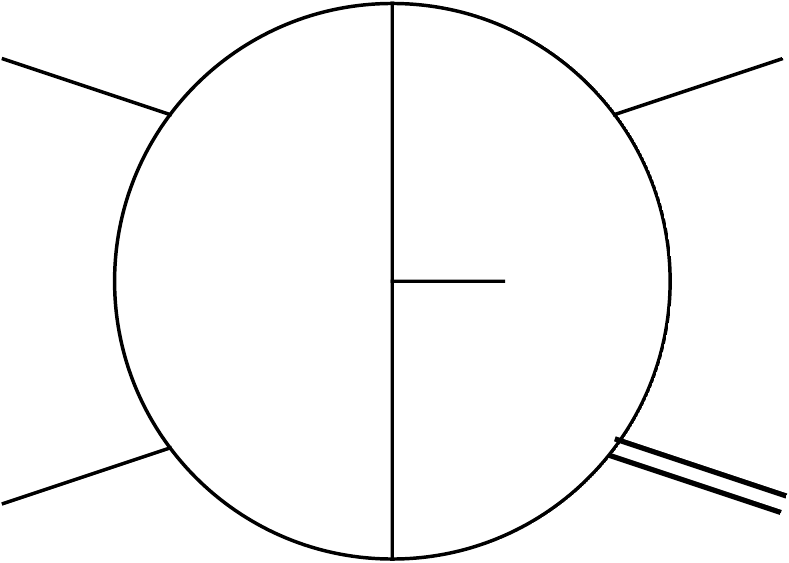}
\caption{Diagrammatic representation of the planar and non-planar families with one external massive leg (double line). In the first row, $P_1$ (left), $P_2$ (middle) and $P_3$ (right) planar families are shown. In the second and third row,
$N_1$ (top left), $N_2$ (top middle), $N_3$ (top right), $N_4$ (bottom left), $N_5$ (bottom right) non-planar families are shown. All internal particles are massless.}
\label{fig:fivepoint}
\end{figure}

Five-point two-loop Master Integrals determine the current frontier. The computation of all planar and non-planar five-point two-loop Master Integrals with massless internal propagators and on-shell light-like external momenta, has been recently completed~\cite{Papadopoulos:2015jft,Gehrmann:2015bfy,Chicherin:2018old,Abreu:2018aqd}. The next step on this path of computing the five-point two-loop Master Integrals would be those with one of the external legs being off-shell. The planar and non-planar topologies corresponding to these Master Integrals are shown in Fig.~\ref{fig:fivepoint}. Based on the Simplified Differential Equations (SDE) approach~\cite{Papadopoulos:2014lla}, we have computed and expressed in terms of Goncharov poly-logarithms, all Master Integrals for the first non-trivial planar family of five-point two-loop Master Integrals with massless internal propagators and one external particle carrying a space- or time-like momentum, $P_1$ in Fig.~\ref{fig:fivepoint},
as well as the full set of planar five-point two-loop massless Master Integrals with light-like external momenta~\cite{Papadopoulos:2015jft}. 
Very recently results on all planar families have been reported in reference~\cite{Abreu:2020jxa}. 
In this paper we present fully analytic results in terms of poly-logarithmic functions for all planar families, based on the Simplified Differential Equations approach. 

In section \ref{sec1}, we  define the 
scattering kinematics and the corresponding integral representations of the Master Integrals and derive the form of the canonical differential equation in the SDE approach.
The derivation of the boundary terms and the solution for all Master Integrals in terms of Goncharov poly-logarithms (GP), is presented in section \ref{sec2}. In section \ref{sec3} we show how to obtain numerical results from our analytic expressions in all kinematical regions.
Finally in section \ref{sec:outlook} we summarize our findings and discuss future applications with emphasis on the computation of the remaining non-planar five-point two-loop Master Integrals.

\section{Planar two-loop five-point Master Integrals with one off-shell leg}
\label{sec1}

There are three families of Master Integrals, labelled as $P_1$, $P_2$ and $P_3$, see Fig.~\ref{fig:fivepoint}, associated to
planar two-loop five-point amplitudes with one off-shell leg.
We adopt the definition of the scattering kinematics following~\cite{Abreu:2020jxa}, where external momenta $q_i,\; i=1\ldots 5$ satisfy $\sum_1^5 q_i=0$, $q_1^2\equiv p_{1s}$, $q_i^2=0,\; i=2\ldots 5$, and the six independent invariants are given by $\{ q_1^2,s_{12},s_{23},s_{34},s_{45},s_{15}\}$, with $s_{ij}:=\left(q_i+q_j\right)^2$.

In the SDE approach~\cite{Papadopoulos:2014lla} the momenta are parametrized by introducing a dimensionless variable $x$, as follows 
\begin{gather}
q_1 \to p_{123}-x p_{12},\; q_2 \to p_4,\; q_3 \to -p_{1234},\; q_4 \to x p_1
\label{eq:momx}
\end{gather}
where the new momenta $p_i,\; i=1\ldots 5$ satisfy now $\sum_1^5p_i=0$, $p_i^2=0,\; i=1\ldots 5$, whereas  $p_{i\ldots j}:=p_i+\ldots +p_j$. The set of independent invariants is given  by $\{ S_{12},S_{23},S_{34},S_{45},S_{51},x\}$, with $S_{ij}:=\left(p_i+p_j\right)^2$. The explicit mapping between the two sets of invariants is given by

\begin{gather}
p_{1s}=(1-x)(S_{45}-S_{12} x),
\;s_{12}=\left(S_{34} - S_{12}(1 - x)\right) x,
\;s_{23}= S_{45},
\;s_{34} = S_{51} x,
\nonumber \\
\;s_{45}=S_{12}x^2,
\;s_{15}=S_{45} + (S_{23} - S_{45}) x
    \label{eq:itatoours}
\end{gather}
and as usual the $x=1$ limit corresponds to the on-shell kinematics.

The corresponding Feynman Integrals are defined through

\begin{gather}
G^{P_1}_{a_1\cdots a_{11}}:=e^{2\gamma_E \epsilon} \int \frac{d^dk_1}{i\pi^{d/2}}\frac{d^dk_2}{i\pi^{d/2}}
\frac{1}{k_1^{2a_1} (k_1 +  q_1)^{2a_2} (k_1 +  q_{12})^{2a_3} (k_1 + q_{123})^{2a_4}} \nonumber\\
\times \frac{1}{ k_2^{2a_5} (k_2 + q_{123})^{2a_6} (k_2 + q_{1234})^{2a_7} (k_1-k_2)^{2a_{8}}(k_1 + q_{1234})^{2a_9}
(k_2 + q_1)^{2a_{10}}(k_2 + q_{12})^{2a_{11}} }
, \label{eq:P1}
\end{gather}

\begin{gather}
G^{P_2}_{a_1\cdots a_{11}}:=e^{2\gamma_E \epsilon} \int \frac{d^dk_1}{i\pi^{d/2}}\frac{d^dk_2}{i\pi^{d/2}}
\frac{1}{k_1^{2a_1} (k_1 - q_{1234})^{2a_2} (k_1 -  q_{234})^{2a_3} (k_1 - q_{34})^{2a_4}} \nonumber\\
\times \frac{1}{k_2^{2a_5} (k_2 - q_{34})^{2a_6} (k_2 - q_4)^{2a_7} 
(k_1 - k_2)^{2a_8} (k_2 - q_{1234})^{2a_{9}} (k_2 - q_{234})^{2a_{10}} (k_1 - q_4)^{2a_{11}}}, \label{eq:P2}
\end{gather}

\begin{gather}
G^{P_3}_{a_1\cdots a_{11}}:=e^{2\gamma_E \epsilon} \int \frac{d^dk_1}{i\pi^{d/2}}\frac{d^dk_2}{i\pi^{d/2}}
\frac{1}{k_1^{2a_1} (k_1 + q_2)^{2a_2} (k_1 +  q_{23})^{2a_3} (k_1 + q_{234})^{2a_4}} \nonumber\\
\times \frac{1}{k_2^{2a_5} (k_2 + q_{234})^{2a_6} (k_2 - q_1)^{2a_7} 
(k_1 - k_2)^{2a_8} (k_1 - q_1)^{2a_{9}} (k_2 + q_2)^{2a_{10}} (k_2 + q_{23})^{2a_{11}}}, \label{eq:P3}
\end{gather}

where $q_{i\ldots j}:=q_i+\ldots +q_j$.

\begin{figure}[h]
\centering
\includegraphics[width=0.3 \linewidth]{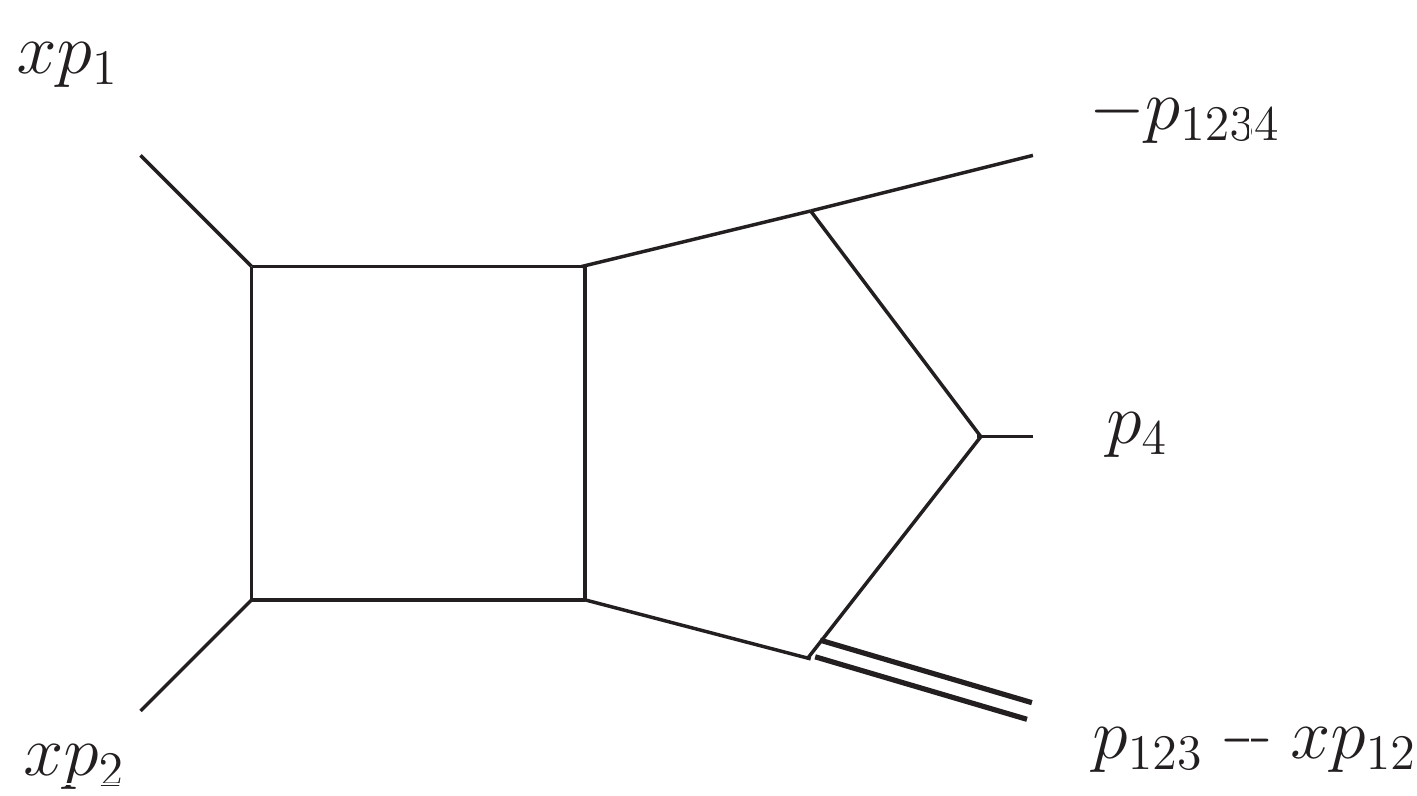}
\includegraphics[width=0.3 \linewidth]{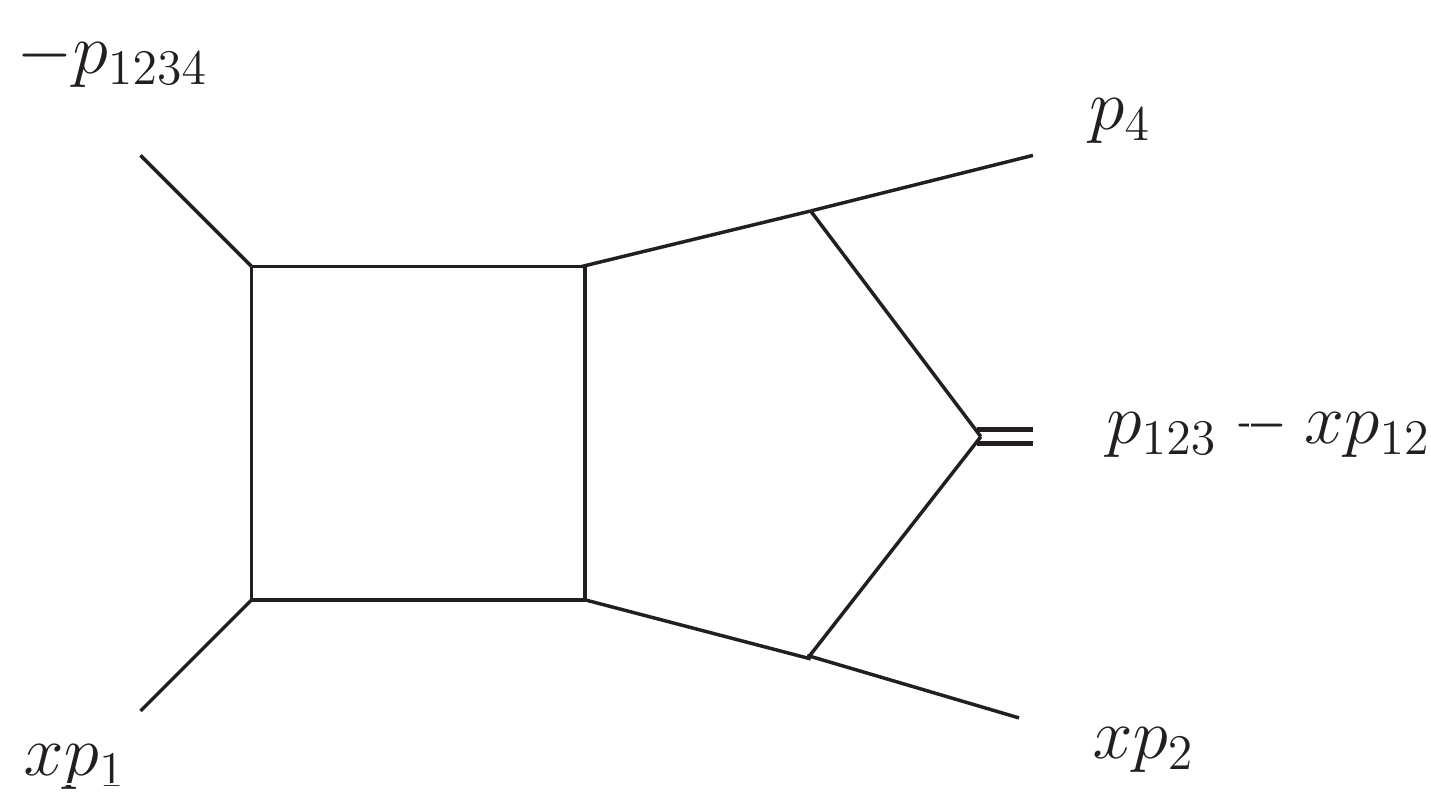}
\includegraphics[width=0.3 \linewidth]{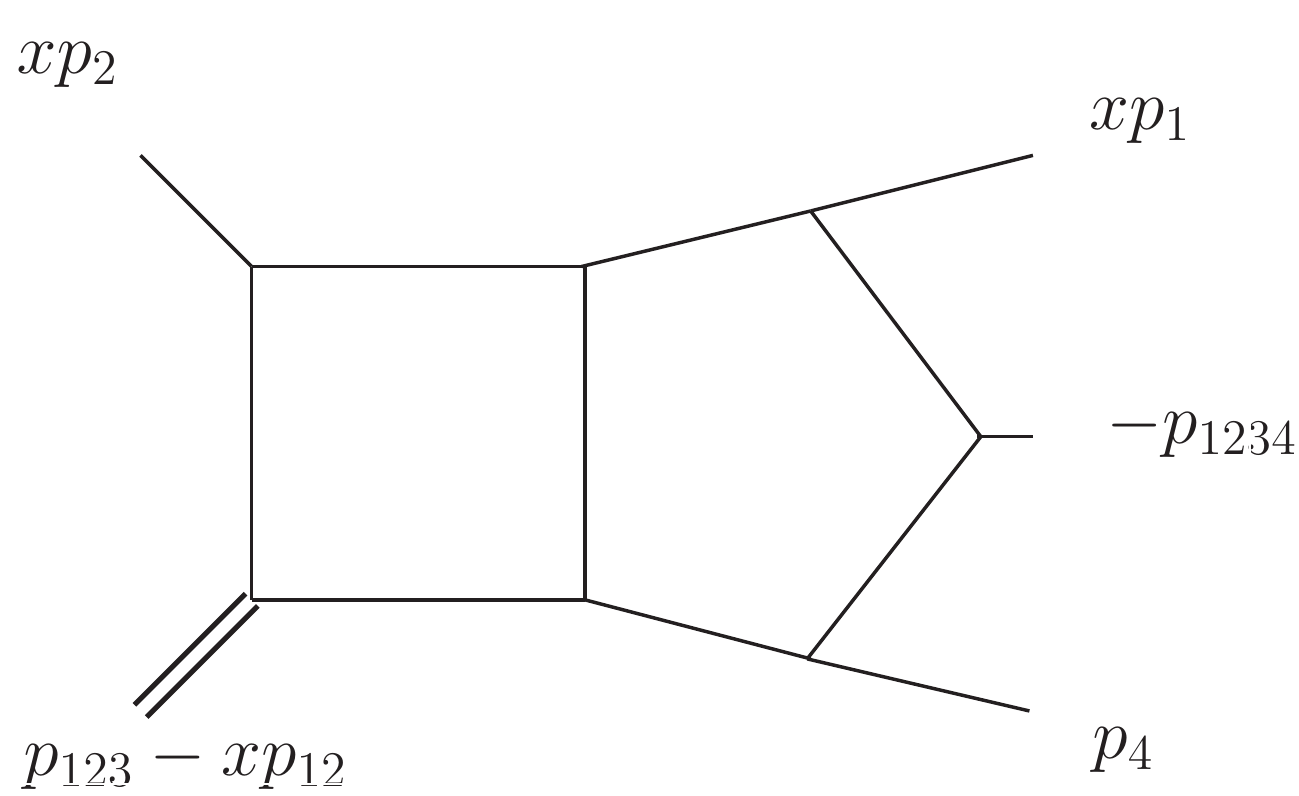}
\caption{The two-loop diagrams representing the top-sector of the planar pentabox family $P_1$, $P_2$ and $P_3$. All external momenta are incoming.}
\label{fig:P1P2P3}
\end{figure}

The $P_1$ family consists of 74 Master integrals. For $P_2$ and $P_3$ the corresponding numbers are 75 and 86. This can easily be verified using standard IBP reduction software, such as {\tt FIRE6}~\cite{Smirnov:2019qkx} and {\tt Kira}~\cite{Maierhoefer:2017hyi,Klappert:2020nbg}. 
The top-sector integrals are shown in Fig.~\ref{fig:P1P2P3}.

\subsection{Canonical basis and Differential Equations}

In order to express all integrals given by Eqs.(\ref{eq:P1} -- \ref{eq:P3}), the easiest way is to define a basis that satisfies a canonical differential equation. By basis we mean a combination of Feynman Integrals with  coefficients depending on the set of invariants and the dimensionality of space-time $d=4-2\epsilon$. Let us assume that such a basis is known, then the DE is written in general as 
\begin{gather}
d\vec g = \epsilon \sum\limits_a {d\log \left( {{W_a}} \right){\tilde{M}_a}\vec g} 
    \label{eq:canonical}
\end{gather}
where $\vec{g}$ represents a vector containing all elements of the canonical basis, $W_a$ are functions of the kinematics and $\tilde{M}_a$ are matrices independent of the kinematical invariants, whose matrix elements are pure rational numbers. Notice that Eq.~(\ref{eq:canonical}) is a multi-variable equation and in the case under consideration the differentiation is understood with respect to the six-dimensional array of independent kinematical invariants, $\{ q_1^2,s_{12},s_{23},s_{34},s_{45},s_{15}\}$. Since $W_a$ are in general algebraic functions of the kinematical invariants a straightforward integration of Eq.~(\ref{eq:canonical}) in terms of generalized poly-logarithms is not an easy task. 

In the SDE approach though, Eq.~(\ref{eq:canonical})  takes the much simpler form 
\begin{gather}
\frac{{d\vec g}}{{dx}} = \epsilon \sum\limits_b {\frac{1}{{x - {l_b}}}{M_b}\vec g} 
    \label{eq:canonicalx}
\end{gather}
where $M_b$ are again rational matrices independent of the kinematics, and the so-called letters, $l_b$, are independent of $x$, depending only on the five invariants, $\left\{ S_{12},S_{23},S_{34},S_{45},S_{51}\right\}$.
Notice that the number of letters in $x$ is generally smaller than the number of letters in Eq.~(\ref{eq:canonical}). Since the Eq.~(\ref{eq:canonicalx}) is a Fuchsian system of ordinary differential equations, it is straightforwardly integrated in terms of Goncharov poly-logarithms, ${\cal G}\left(l_1,l_2,\ldots;x\right)$.

Over the last years much effort has been devoted to construct the canonical basis, or at least an educated guess of it, and then verify the form of Eq.~(\ref{eq:canonical}) through standard differentiation and IBP reduction. We refer to section 4 of reference~\cite{Abreu:2020jxa} for a thorough discussion of relevant work in the literature. In principle the knowledge of the canonical basis is enough within the SDE approach to derive the form of the corresponding canonical differential equation, Eq.~(\ref{eq:canonicalx}), by explicitly differentiating with respect to $x$ and using IBP identities to express the resulting combinations of Feynman integrals in terms of basis elements. In fact, as we will show in section~\ref{sec:SDE}, since the matrices entering in Eq.~(\ref{eq:canonicalx}) are independent of the kinematics, one can use solutions of IBP identities derived by assigning integer values to the kinematics, except $x$. Using nowadays packages such as \texttt{FIRE6} and \texttt{Kira-2.0} the above-mentioned IBP-reduction becomes a computationally trivial exercise. Notice that there is no need to use rational reconstruction methods, as far as the derivation of Eq.~(\ref{eq:canonicalx}) is concerned. 

\subsection{The Simplified Differential Equations}
\label{sec:SDE}

\begin{figure}[t!]
\centering
\includegraphics[width=0.4\linewidth]{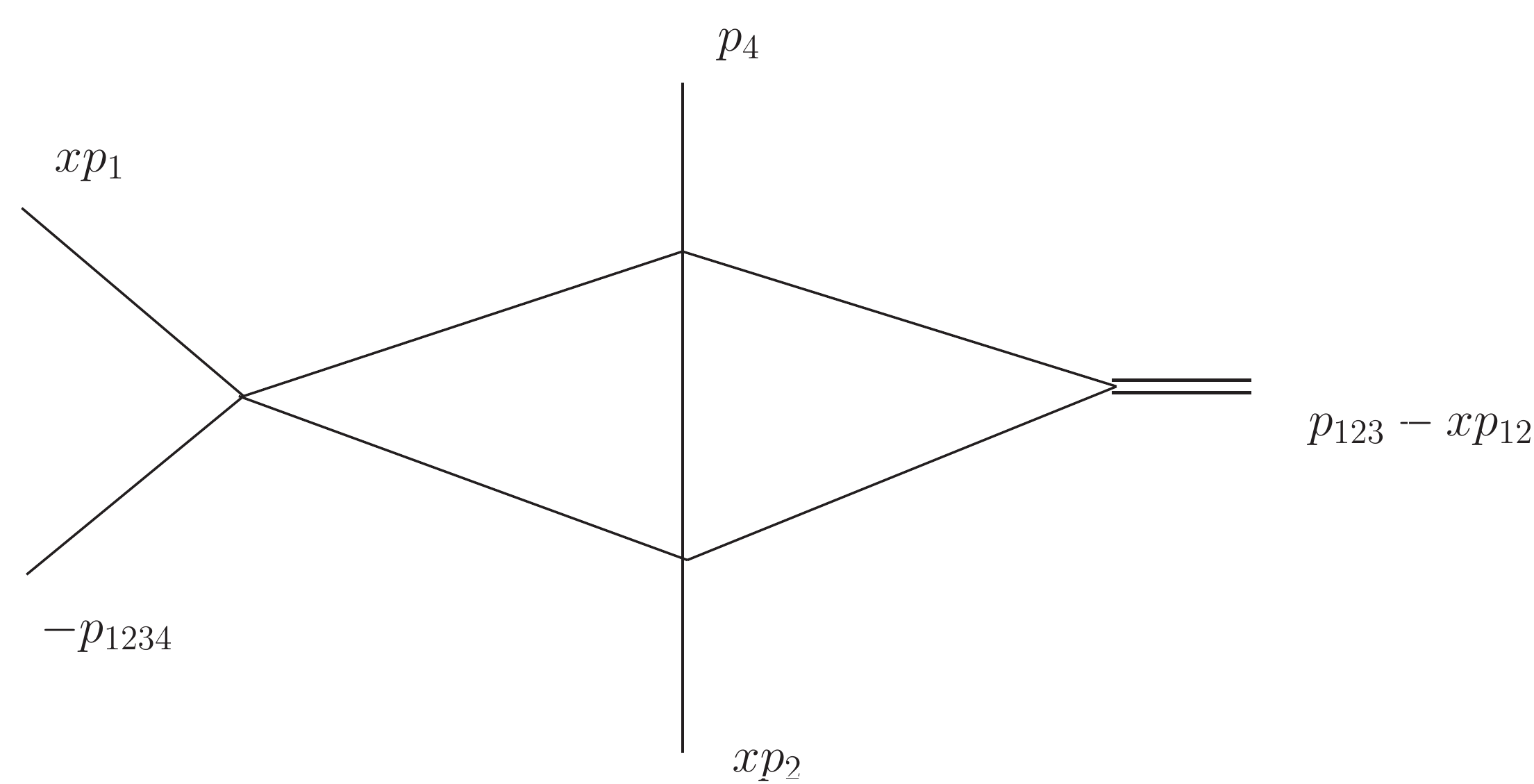}
\caption{The two-loop diagram representing the decoupling basis element.}
\label{fig:p246}
\end{figure}

Knowing from reference~\cite{Abreu:2020jxa}, the explicit form of the matrices ${\tilde M}_a$ and of the letters $W_a$ in terms of the variables $p_{1s},s_{12},s_{23},s_{34},s_{45},s_{15}$ ($p_{1s}\equiv q_1^2$), in Eq.~(\ref{eq:canonical}), we simply derive the data needed in Eq.~(\ref{eq:canonicalx}), based on the following identity,
\begin{gather}
    \sum\limits_a {\frac{{d\log \left( {{W_a}} \right)}}{{dx}}{{\tilde M}_a}}  \equiv \sum\limits_b {\frac{1}{{x - {l_b}}}{M_b}} 
    \label{eq:identity}
\end{gather}
making use of Eq.~(\ref{eq:itatoours}).
For $P_2$ and $P_3$ families Eq.~(\ref{eq:identity}) is applicable after eliminating a special basis element whose leading singularity is proportional to a non-rationalizable square root in terms of $x$. The corresponding integral is shown in Fig.~\ref{fig:p246} and it is the same for the two families. Its expression in terms of poly-logarithmic functions is already known from the double-box families with two off-shell legs\footnote{The basis element is numbered as 46 in the $P_2$ family and 53 in the  $P_3$ family and is given in terms of the double-box $P_{23}$ family variables~\cite{Papadopoulos:2014hla}, whose expression in terms of the pentabox kinematics is given in the ancillary files {\tt anc/P2/Letters} and/or {\tt anc/P3/Letters}.}.
Since our task is to evaluate all basis elements up to ${\cal O}(\epsilon^4)$ and since the basis element expansion of the above integral starts at ${\cal O}(\epsilon^4)$, it effectively decouples from the differential equation Eq.~(\ref{eq:canonicalx}). 
Nevertheless,  for the sake of completeness, notice that by parametrizing the momenta in terms of a new variable $y$, 
\begin{gather}
q_1 \to p^\prime_{123}-y p^\prime_{12},\; q_2 \to y p^\prime_1,\; q_3 \to p^\prime_4 ,\; q_4 \to -p^\prime_{1234}
\label{eq:momy}
\end{gather}
resulting into
\begin{gather}
p_{1s}=(1 - y) (S^\prime_{45} - S^\prime_{12} y),
\;s_{12}=S^\prime_{45} - (S^\prime_{12} + S^\prime_{23}) y,
\;s_{23}=\left(S^\prime_{34} - S^\prime_{12} (1 - y)\right) y,
\nonumber \\
\;s_{34} = S^\prime_{45},
\;s_{45}= -(S^\prime_{12} - S^\prime_{34} + S^\prime_{51}) y,
\;s_{15}= S^\prime_{45} + S^\prime_{23} y
    \label{eq:itatooursy}
\end{gather}
we can effectively rationalize the square roots related to the special basis element.
By using then both variables, Eq.~(\ref{eq:identity}) can be generalized in the form

\begin{gather}
d\vec g = \epsilon \left[ {\sum\limits_b {d\log \left( {x - {l_b}} \right){M_b}}  + \sum\limits_c {d\log \left( {y - {\bar{l}_c}} \right){\bar{M}_c}}  + d\log \left( {{W_{58}}} \right){{\tilde M}_{58}}} \right]\vec g.
    \label{eq:canonicalxy}
\end{gather}
This is achieved because all letters $W_a$ in Eq.~(\ref{eq:canonical}), except $W_{58}$, are linear functions only of $x$ or $y$. 

The full list of letters $l_b$ and matrices $M_b$, Eq.~(\ref{eq:canonicalx}), for the $P_1$ family, are given in the ancillary files provided 
as {\textbf{anc/P1/Letters}}, {\textbf{anc/P1/Matrices}} and in the same notation for $P_2$ and $P_3$ families. 
Moreover, the explicit expressions of the letters for all families is given in the Appendix~\ref{App:A}.


\section{Boundary Conditions and Analytic Expressions}
\label{sec2}

The solution of Eq.~(\ref{eq:canonicalx}) up to order $\mathcal{O}\left(\epsilon^4\right)$ can be written as follows:
\begin{align}
   \label{eq:solution}
   \textbf{g}&= \epsilon^0 \textbf{b}^{(0)}_{0} + \epsilon \bigg(\sum\mathcal{G}_{a}\textbf{M}_{a}\textbf{b}^{(0)}_{0}+\textbf{b}^{(1)}_{0}\bigg) \nonumber \\
   &+ \epsilon^2 \bigg(\sum\mathcal{G}_{ab}\textbf{M}_{a}\textbf{M}_{b}\textbf{b}^{(0)}_{0}+\sum\mathcal{G}_{a}\textbf{M}_{a}\textbf{b}^{(1)}_{0}+\textbf{b}^{(2)}_{0}\bigg) \nonumber \\
   &+ \epsilon^3 \bigg(\sum\mathcal{G}_{abc}\textbf{M}_{a}\textbf{M}_{b}\textbf{M}_{c}\textbf{b}^{(0)}_{0}+\sum\mathcal{G}_{ab}\textbf{M}_{a}\textbf{M}_{b}\textbf{b}^{(1)}_{0}+\sum\mathcal{G}_{a}\textbf{M}_{a}\textbf{b}^{(2)}_{0}+\textbf{b}^{(3)}_{0}\bigg) \nonumber \\
   &+ \epsilon^4 \bigg(\sum\mathcal{G}_{abcd}\textbf{M}_{a}\textbf{M}_{b}\textbf{M}_{c}\textbf{M}_{d}\textbf{b}^{(0)}_{0}+\sum\mathcal{G}_{abc}\textbf{M}_{a}\textbf{M}_{b}\textbf{M}_{c}\textbf{b}^{(1)}_{0}\nonumber \\
   &+ \sum\mathcal{G}_{ab}\textbf{M}_{a}\textbf{M}_{b}\textbf{b}^{(2)}_{0}+\sum\mathcal{G}_{a}\textbf{M}_{a}\textbf{b}^{(3)}_{0}+\textbf{b}^{(4)}_{0}\bigg)
   \\
   \mathcal{G}_{ab\ldots}&:= \mathcal{G}(l_a,l_b,\ldots;x) \nonumber
\end{align}
where \textbf{g} and \textbf{M} represent $\vec{g}$ and $M$ appearing in Eq.~(\ref{eq:canonicalx}) and 
$\textbf{b}^{(i)}_{0}$ are the boundary values of the basis elements in the limit $x\to 0$ (see Eq.(3.6) of reference~\cite{Papadopoulos:2015jft}) at order $\epsilon^i,\;i=0\ldots 4$. 
In the above equation $\mathcal{G}(l_a,l_b,\ldots;x)$ stands for Goncharov polylogarithms. Since all the data of the above equation, namely the letters $l_a,l_b,\ldots$ and the matrices $\textbf{M}_a,\textbf{M}_b,\ldots$ are already given, the only remaining task is the computation of the boundary values, $\textbf{b}^{(i)}_{0}$, in terms of poly-logarithmic functions. 

To derive the $x\to 0$ limit of basis elements  we first exploit the canonical differential equation in $x$, Eq.~(\ref{eq:canonical}), which in the limit takes the form 
\begin{gather}
\frac{{d\vec g}}{{dx}} = \epsilon  {\frac{1}{x}{M_0}\vec g}+\mathcal{O}(x^0) 
    \label{eq:canonicalx0}
\end{gather}
with the solution ($\textbf{b}:=\sum_{i=0}^4\epsilon^i\textbf{b}_0^{(i)}$)
\begin{gather}
\textbf{g}_0=\textbf{S}e^{\epsilon \log(x) \textbf{D}}\textbf{S}^{-1} \textbf{b}   
 \label{eq:solutionx0}
\end{gather}
and the matrices \textbf{S} and \textbf{D} are obtained through Jordan decomposition of the $\textbf{M}_0$ matrix, $\textbf{M}_0=\textbf{S}\textbf{D}\textbf{S}^{-1}$.
We call the matrix $\textbf{R}_0=\textbf{S}e^{\epsilon \log(x) \textbf{D}}\textbf{S}^{-1}$, the resummed matrix at $x=0$. Since the biggest Jordan block of it has dimension two, it can be written in the form
\begin{equation}
{{\bf{R}}_0} = \sum\limits_i {{x^{{n_i}\varepsilon }}\left( {{{\bf{R}}_{0i}} + \varepsilon \log \left( x \right){{\bf{R}}_{0i0}}} \right)} 
 \label{eq:resum0}
\end{equation}
with $\textbf{R}_{0i}$ and $\textbf{R}_{0i0}$ matrices of rational numbers and the exponents $n_i$ are the eigenvalues of the matrix \textbf{D} (equivalently $\textbf{M}_0$).

On the other hand through IBP reduction the elements of the canonical basis can be related to a set of Master Integrals,
\begin{equation}
    \textbf{g}=\textbf{T}\textbf{G}.
    \label{eq:gtomasters}
\end{equation}
The list of Feynman Integrals \textbf{G} chosen as Master Integrals in the IBP reduction as well as the expression of the basis elements in terms of Feynman Integrals for all families is given in the corresponding ancillary files, \textbf{anc/P1...P3/Masters} and \textbf{anc/P1...P3/Basis}.

We have used the expansion by regions techniques~\cite{Jantzen:2012mw} in order to write each Master Integral in the form of a sum over region-integrals, 
\begin{equation}
{G_i}\mathop  = \limits_{x \to 0} \sum\limits_j {{x^{{b_j} + {a_j}\varepsilon }}G_i^{\left( j \right)}} 
\label{eq:regions}
\end{equation}
with $a_j$ and $b_j$ being integers, by making use of the \texttt{FIESTA4}~\cite{Smirnov:2015mct} public code.
Combining Eqs.~(\ref{eq:solutionx0}) and (\ref{eq:gtomasters}) we get
\begin{equation}
{\bf g_0} : = {{\bf{R}}_0}{\bf{b}} = \mathop {\lim }\limits_{x \to 0} \left. {{\bf{TG}}} \right|_{{\cal O}(x^{0+a_j\epsilon})}
\label{eq:boundary}
\end{equation}
where, since the dependence of the left-hand side on $x$ is only through Eq.~(\ref{eq:resum0}), in the right-hand side, except for the terms of the form  $x^{a_j\epsilon}$ arising from Eq.~(\ref{eq:regions}), we expand around $x=0$, keeping only terms of order $x^0$. Notice also, that the left-hand side of the equation contains the boundary values of the basis elements that are pure functions of the underlying kinematics $\vec{S}:=\{ S_{12},S_{23},S_{34},S_{45},S_{51}\}$ whereas in the right hand side the matrix \textbf{T} is an algebraic function of $\vec{S}$. The consistency of Eq.~(\ref{eq:boundary}) implies that the right-hand side should also be a pure function of $\vec{S}$. Therefore, in order to determine the matrix \textbf{T} entering in Eq.~(\ref{eq:boundary}), we can employ solutions of IBP identities using numerical, actually integer values for $\vec{S}$, keeping $x$ and $d$ in a symbolic form. This results to a significant reduction in complexity and CPU time, taking into account that there are several basis elements in $ \textbf{g}$, that are given in terms of Baikov polynomials~\cite{Abreu:2020jxa}, $\mu_{11}, \mu_{12}, \mu_{22}$, which when expressed in terms of inverse propagators, contain Feynman Integrals with up to fourth powers of irreducible inverse propagators. 

It turns out that Eq.~(\ref{eq:boundary}) is a powerful framework allowing to determine all boundary constants \textbf{b}. First of all in the case that the left-hand side contains a logarithmic term in $x$, a set of linear relations between elements of the array \textbf{b} are obtained by setting the coefficient of  $x^{a_j\epsilon}\log \left( x \right)$ terms to zero. Secondly, powers of $x^{a_j \epsilon}$ that appear only in the left-hand side, also produce relations among elements of \textbf{b}, by setting their coefficients to zero. These two sets of relations account for the determination of a significant part of the components of the boundary array. The last set of equations requires the determination of some regions of Master Integrals ${G_i^{\left( j \right)}}$ in the right-hand side of Eq.~(\ref{eq:boundary}). Expressions of the integrands of these region-integrals ${G_i^{\left( j \right)}}$, in terms of Feynman parameters are obtained using \texttt{SDExpandAsy} in \texttt{FIESTA4}~\cite{Smirnov:2015mct}. Their calculation is straightforwardly achieved, either by direct integration in Feynman-parameter space and then by using \texttt{HypExp}~\cite{Huber:2005yg,Huber:2007dx} to expand the resulting $_2{F_1}$ hypergeometric functions, or in a very few cases, by Mellin-Barnes techniques using the \texttt{MB}~\cite{Czakon:2005rk,mbtools},  \texttt{MBSums}~\cite{Ochman:2015fho} and \texttt{XSummer}~\cite{Moch:2005uc} packages\footnote{The in-house \texttt{Mathematica} package \texttt{Gsuite}, that automatically process the \texttt{MBSums} output through \texttt{XSummer}, written by A.~Kardos, is used.}. 
The above described procedure leads to fully analytic expressions of the region-integrals ${G_i^{\left( j \right)}}$ in terms of $\vec{S}$. Their general structure is given in terms of logarithmic and poly-logarithmic functions of $\vec{S}$, as well as rational factors depending on $\vec{S}$. When implementing ${G_i^{\left( j \right)}}$ in the right-hand side of Eq.~(\ref{eq:boundary}), their rational factors are evaluated on the same numerical values for $\vec{S}$ as in the determination of the matrix \textbf{T}, described in the previous paragraph. We have of course verified that the results are indeed independent of the choice of the specific numerical assignment for the variables in $\vec{S}$.  
All the boundary values $ \textbf{b}$, are analytically expressed in terms of poly-logarithmic functions, namely logarithms and Goncharov poly-logarithms depending on the reduced kinematical variables $\vec{S}$, and are manifestly pure functions. Although the above described method is general and straightforward, in practice many of the components of $\textbf{b}$ have been obtained by exploiting the known representations of the elements of the canonical basis as given in the double-box families~\cite{Papadopoulos:2014hla,Henn:2014lfa}. Boundary terms  $\textbf{b}$ and basis elements  $\textbf{g}$, expressed in terms of poly-logarithmic functions can be found in the ancillary files \textbf{anc/P1...P3/Boundaries} and \textbf{anc/P1...P3/Results}, respectively.  

\section{Numerical Results and Validation}
\label{sec3}

In order to numerically evaluate the solution given in Eq.~(\ref{eq:solution}), Goncharov poly-logarithms up to weight 4 need to be computed. To understand the complexity of the expressions at hand, we present in Table~\ref{tab:polylog}, the number of poly-logarithmic functions entering in the solution. In parenthesis we give the corresponding number for the non-zero top-sector basis elements. The weight W=$1\ldots 4$ is identified as the number of letters $l_a$ in GP $\mathcal{G}(l_a,\ldots;x)$. 

\begin{table}[ht]
    \centering
\begin{tabular}{|c|c|c|c|c|}
\hline
 Family     & W=1 & W=2 & W=3 & W=4 
 \\
 \hline
$P_1$ ($g_{72}$)      & 17 (14)   & 116 (95) & 690 (551) & 2740 (2066) 
\\
\hline
$P_2$ ($g_{73}$)      & 25 (14)   & 170 (140) & 1330 (1061) & 4950 (3734) 
\\
\hline
$P_3$ ($g_{84}$)     & 22 (12)  & 132 (90) & 1196 (692) & 4566 (2488) 
\\
\hline
\end{tabular}
\caption{Number of GP entering in the solution, as explained in the text.}
\label{tab:polylog}
\end{table}

The computation of GPs is performed using their implementation in \texttt{GiNaC}~\cite{ginac}. 
This implementation is capable to evaluate the GPs at an arbitrary precision. The computational cost to numerically evaluate a GP function, depends of course on the number of significant digits required as well as on their weight and finally on their structure, namely how many of its letters, Eq.~(\ref{eq:solution}), satisfy ${l_a} \in \left[ {0,x} \right]$.
We refer to reference~\cite{Vollinga:2004sn} for more details.

For the following Euclidean point 
\begin{equation}
    S_{12}\to -2,S_{23}\to -3,S_{34}\to -5,S_{45}\to -7,S_{51}\to -11,x\to \frac{1}{4}
\end{equation}
all GP functions with real letters are real, namely no letter is in $\left[ {0,x} \right]$,  and moreover the boundary terms are by construction all real.
The result is given in Table~\ref{tab:pointEx}
\begin{table}[h!]
    \centering
    \begin{tabular}{|c|c|c|}
    \hline
      $P_1$  & $g_{72}$ & 
      \parbox[c][2.6cm][c]{7cm}{ 
      $\epsilon^0$: 3/2
      \\
      $\epsilon^1$:
      -2.2514604753379400332169314784961
      \\
      $\epsilon^2$:
      -17.910593443812320786572184851867
      \\
      $\epsilon^3$:
      -26.429770706459534336624681550003
      \\
      $\epsilon^4$:
      21.437938934510558345847354772412
      }
      \\
    \hline 
     $P_2$ & $g_{73}$ & \parbox[c][2.2cm][c]{7cm}{
     $\epsilon^1$: 
     2.8124788185742741402751457351382
      \\
      $\epsilon^2$:
      5.4813042746593704203645729908938
      \\
      $\epsilon^3$:
      11.590234540689191439870956817546
      \\
      $\epsilon^4$:
      -5.9962816226829136730734255754596
      }
     \\ 
    \hline
    $P_3$ & $g_{84}$ & \parbox[c][2.6cm][c]{7cm}{ 
    $\epsilon^0$:
    1/2 \\
    $\epsilon^1$:
    3.2780415861887284967738281876762
      \\
      $\epsilon^2$:
      0.11455863130537720411162743574627
       \\
      $\epsilon^3$:
      -16.979642659429606120982671925458
      \\
      $\epsilon^4$:
      -48.101985355625914648042310964575
      } 
    \\
    \hline
    \end{tabular}
    \caption{Numerical results for the non-zero top sector element of each family with 32 significant digits.}
    \label{tab:pointEx}
\end{table}
 with timings, running the \texttt{GiNaC} Interactive Shell \texttt{ginsh}, given by $1.9$, $3.3$, and $2$ seconds for $P_1$, $P_2$ and $P_3$ respectively and for a precision of 32 significant digits. As can be seen from Table~\ref{tab:polylog}, the number of $W=4$ GPs of the top-sector element is more than 50\% of the total number of $W=4$ GPs in each family. Taking into account that the vast majority of CPU time is spent in the evaluation of $W=4$ GPs, the CPU time to compute the full list of basis elements in a family, is of the same order of magnitude as the computation of its top-sector element. 
 
In order to obtain numerical results for scattering kinematics,
we need to properly analytically continue the GPs and logarithms involved in our solution, Eq.~(\ref{eq:solution}). The easiest way is to determine for each physical point under consideration, the real parameters $\delta_{ij}$ and $\delta_{x}$ so that the substitution, $S_{ij}\to S_{ij} + i\delta_{ij}\eta, $ $x\to x + i\delta_{x}\eta$, $\eta\to 0$, of the variables used in our solution, properly accounts for the analytic continuation.  As detailed in references~\cite{Papadopoulos:2014hla,Papadopoulos:2015jft}, $\delta_{ij}$ and $\delta_{x}$ should  satisfy analyticity constraints stemming (a) from the second graph polynomial ${\cal F}$ of the top-sector Feynman integral and (b) from the representation of the 
one-scale integrals in terms of the variables $x$ and $S_{ij}$.

First notice that in our case,
when the kinematic variables $p_{1s}$ and $s_{ij}$, acquire an infinitesimal imaginary part~\cite{Gehrmann:2002zr}, namely $s_{ij}\to s_{ij} + i\eta,$ $p_{1s}\to p_{1s} + i\eta$, with $\eta\to 0$, the sign of the imaginary part of the second graph polynomial ${\cal F}$ is consistent with the $i0$ prescription of the Feynman propagator. 
We solve Eq.~(\ref{eq:itatoours}) in terms of ${\bar S}_{ij}:=S_{ij} + i\delta_{ij}\eta$ and ${\bar x}:= x + i\delta_{x}\eta$, using as input ${\bar s}_{ij}:= s_{ij} + i\eta$ and ${\bar p}_{1s}:=p_{1s} + i\eta$ and 
we find two solutions for  ${\bar S}_{ij}$ and ${\bar x}$. The numerical value of $\eta$ 
is chosen to match the required precision of the results.
We then check if the obtained solutions satisfy the constraints stemming from the one-scale integrals. These integrals are proportional to $(-s_{ij})^{n\epsilon}$, $(-p_{1s})^{n\epsilon}$, $n=-1,-2$, and their expressions in terms $S_{ij}$ and $x$ used to obtain the analytic solution in Eq.~(\ref{eq:solution}), are given as follows:
\begin{align}
\centering
   &(-s_{34})^{-\epsilon} = (-S_{51})^{-\epsilon} x^{-\epsilon}\nonumber\\
   &(-s_{45})^{-\epsilon} = (-S_{12})^{-\epsilon} x^{-2\epsilon}\nonumber\\
   &(-s_{15})^{-\epsilon} = (-S_{45})^{-\epsilon} \bigg(1-\frac{S_{45}-S_{23}}{S_{45}}x\bigg)^{-\epsilon}\nonumber\\
   &(-p_{1s})^{-\epsilon} = (1-x)^{-\epsilon} (-S_{45})^{-\epsilon} \bigg(1-\frac{S_{12}}{S_{45}}x\bigg)^{-\epsilon}\nonumber\\
   &(-s_{12})^{-\epsilon} = x^{-\epsilon} (S_{12}-S_{34})^{-\epsilon} \bigg(1-\frac{S_{12}}{S_{12}-S_{34}}x\bigg)^{-\epsilon}.
\label{eq:onescale}
\end{align}
In practice any solution that satisfies the above equations, with $s_{ij} \to {\bar s}_{ij}$, $p_{1s}\to{\bar p}_{1s}$ and $S_{ij}\to {\bar S}_{ij}$, $x\to {\bar x}$, can be used for the numerical evaluation of the basis elements. We have found that, for all physical points used in the numerical evaluation of the basis elements below, at least one of the two solutions of Eq.~(\ref{eq:itatoours}) is consistent with Eq.~(\ref{eq:onescale}). 
Finally, when Goncharov poly-logarithmic functions ${\cal G}(l_a, l_b,\ldots;x)$, with $0 < {l_a},{l_b}, \ldots  < x$, need to be evaluated, we have  to determine the infinitesimal part of the letters $l_a$ and $x$. This is easily achieved through the explicit expression of the letters in terms of the variables $S_{ij}$, given in Appendix~\ref{App:A} and the solution $S_{ij}\to S_{ij} + i\delta_{ij}\eta$, $x\to x + i\delta_{x}\eta$, as discussed above.
 

In Table~\ref{tab:point-phys-1} we present results for all non-zero top sector elements at $W=4$, for the first physical point provided in reference~\cite{Abreu:2020jxa}, namely
\begin{equation}
   s_{12}\to -\frac{22}{5},\, s_{15}\to \frac{249}{50},\, s_{23}\to \frac{241}{25},\, s_{34}\to -\frac{377}{100},\, s_{45}\to \frac{13}{50},\, \text{p1s}\to \frac{137}{50}.
\end{equation}
In this case no letter lies in the interval $\left[ {0,x} \right]$.
The timings, running the \texttt{GiNaC} Interactive Shell \texttt{ginsh}, are 5.95 (2.33), 11.98 (4.94) and 8.49 (3.32) seconds for $P_1$, $P_2$ and $P_3$ respectively, for $N_{digits}=32\;(16)$.
\begin{table}[h!]
    \centering
    \begin{tabular}{|c|c|c|}
    \hline
      $P_1$  & $g_{72}$ &
      \parbox[c][1cm][c]{7cm}{
      29.802763651793108812023893217593
      \\+$i$ 273.86627846266515113913295225572}
      \\
    \hline
     mzz & $I_3$ & 
     \parbox[c][1cm][c]{7cm}{
      29.802763651793108812023893217593
      \\+$i$ 273.86627846266515113913295225572}
      \\
    \hline
    \hline
      $P_2$  & $g_{73}$ & 
      \parbox[c][1cm][c]{7cm}{44.162165744735300867233118554183
      \\-$i$ 46.218746133850339969944403077557}
      \\
    \hline
     zmz & $I_3$ & \parbox[c][1cm][c]{7cm}{44.162165744735300867233118554183
     \\-$i$ 46.218746133850339969944403077557}
     \\
    \hline
    \hline
      $P_3$  & $g_{84}$ & \parbox[c][1cm][c]{7cm}{11.908529680841593329567378444341
      \\-$i$ 143.83838235097336513553728991658} 
      \\
    \hline
     zzz & $I_3$ & \parbox[c][1cm][c]{7cm}{11.908529680841593329567378444341
     \\-$i$ 143.83838235097336513553728991658} 
     \\
    \hline
    \end{tabular}
    \caption{Numerical results for the non-zero top sector element of each family at weight 4 with 32 significant digits. The notation $I_{i}$ is used in accordance with Table 2 of~\cite{Abreu:2020jxa}.}
    \label{tab:point-phys-1}
\end{table}

For the other physical points, beyond  the first one, the number of letters in $\left[ {0,x} \right]$ is not anymore zero. Since the numerical evaluation relies on the algorithm described in reference~\cite{Vollinga:2004sn}, the running time used by {\tt GiNaC} to compute a
Goncharov poly-logarithmic function, ${\cal G}(l_a, l_b,\ldots;x)$, with $0 < {l_a},{l_b}, \ldots  < x$ , is significantly increasing with the number of letters in $\left[ {0,x} \right]$. As a consequence the CPU time to compute the top-sector basis elements at $W=4$  is also increasing, up to two orders of magnitude, with the last physical point being the worst case, as for this point the number of letters in $\left[ {0,x} \right]$ amounts to 19 out of a total of 24 letters involved in the non-zero top-sector basis elements. It is therefore worthwhile to thoroughly investigate the structure of the analytic result, with the aim to provide alternative representations in terms of Goncharov poly-logarithmic functions that are manifestly real-valued and thus much faster to compute. Notice that, from the structure of the analytic representation studied in this paper (see for instance Table~\ref{tab:polylog}), the computational time is entirely determined by the $W=4$ functions. Therefore, as experience shows~\cite{Abreu:2020xvt,Chicherin:2020oor,Gehrmann:2018yef,Papadopoulos:2019iam}, the use of one-dimensional integral representations at $W=4$, may lead to a  significant reduction in CPU time. We intend to devote  a forthcoming publication to address in detail all these issues. 

We have also compared our results for all families, all basis elements and all physical points with those of reference~\cite{Abreu:2020jxa} and found perfect agreement to the precision used, ($N_{digits}=16,32$). We also checked our results, not only at the level of basis elements but also at the level of Master Integrals, against \texttt{FIESTA4}~\cite{Smirnov:2015mct} and found agreement within the numerical integration errors provided by it. 

\section{Conclusions and Outlook}
\label{sec:outlook}

In this paper we have presented analytic expressions in terms of poly-logarithmic functions, Goncharov Polylogarithms, of all planar two-loop five-point integrals with a massive external leg. This has been achieved by using the Simplified Differential Equations approach and the data for the canonical basis provided in reference~\cite{Abreu:2020jxa}. Moreover, the necessary boundary values of all basis elements have been computed, based mainly on the form of the canonical differential equation, Eq.~(\ref{eq:canonicalx}) and, in few cases, on the expansion by regions approach. The ability to straightforwardly compute the boundary values at $x=0$ and to even more straightforwardly  express the solution in terms of Goncharov Polylogarithms, is based on the unique property of the SDE approach that the scattering kinematics is effectively simplified and rationalized with respect to $x$, in noticeable contradistinction with the standard differential equation approach, where such an analytic realisation of the solution is prohibitively difficult.   

Obviously, the next step, is to extend the work of this paper in the case of the remaining five non-planar families, shown in Fig.~\ref{fig:fivepoint}. Since on top of the planar penta-box families presented in this paper, we have already computed the pure-function solutions in SDE approach, for all double-box families, planar and non-planar, with up to two external massive legs, we expect that the construction of the canonical basis of the few remaining non-planar Master Integrals will be plausible in the near future. Having the corresponding equation, Eq.~(\ref{eq:canonicalx}), for the non-planar families, it should be straightforward to extend the work of this paper and to complete the full list of two-loop five-point Feynman Integrals with one massive external leg. We remind that within the SDE approach, having the analytic representations of two-loop five-point Master Integrals with one massive external leg in terms of Goncharov poly-logarithmic functions, allows also to straightforwardly obtain  the result for massless external legs in terms of Goncharov poly-logarithmic functions, by taking the limit $x=1$~\cite{Papadopoulos:2015jft,Papadopoulos:2019iam} and making use of the resummed matrix corresponding to $l_b=1$ term in Eq.~(\ref{eq:canonicalx}). In summary, when this next step is completed, a library of all two-loop Master Integrals with internal massless particles and up to five (four) external legs, among which one (two) massive legs will be provided: this will constitute a significant milestone towards the knowledge of the full basis of two-loop Feynman Integrals. 

We have also shown how to obtain numerical results for all kinematic configurations, including Euclidean and physical regions. With regard to the expected progress in the calculation of $2\to 3$ scattering process~\cite{Abreu:2018jgq,Hartanto:2019uvl,Abreu:2020xvt}, it would be desirable to adapt our results in different kinematic regions, using for instance fibration-basis techniques~\cite{Panzer:2014caa,Duhr:2019tlz}. We postpone the analysis of the effectiveness of the numerical computation to a forthcoming publication.

\acknowledgments
We would like to  express our gratitude to Chris Wever for many fruitful discussions and assistance with Mellin-Barnes approach, as well as to  Adam Kardos for providing us with the in-house package \texttt{Gsuite}. We also thank Alexander Smirnov, for his valuable assistance in running \texttt{FIRE6} and Johann Usovitsch for advising us on the use of \texttt{Kira-2.0}.
C.G.P. would like also to thank CERN Theory Department for their kind hospitality, during which part of the work has been done.

\noindent This research is co-financed by Greece and the European Union (European Social Fund- ESF)
through the Operational Program Human Resources Development, Education and Lifelong
Learning 2014 - 2020 in the context of the project "Higher order corrections in QCD with applications to High Energy experiments at LHC" -MIS 5047812.

\appendix
\section{The alphabet in \textit{x}}
\label{App:A}

The alphabet for the three planar families considered in this paper consists of 32 letters in total, namely
\begin{align}
    &l_1\to 0,l_2\to 1,l_3\to \frac{S_{12}+S_{23}}{S_{12}},l_4\to 1-\frac{S_{34}}{S_{12}},l_5\to \frac{S_{45}}{S_{12}},l_6\to -\frac{S_{45}}{S_{23}-S_{45}}, \nonumber \\
    &l_7\to \frac{S_{45}-S_{23}}{S_{12}},l_8\to \frac{S_{45}}{S_{34}+S_{45}},l_9\to -\frac{S_{51}}{S_{12}},l_{10}\to \frac{S_{12}-S_{34}+S_{51}}{S_{12}},\nonumber \\
    &l_{11}\to \frac{S_{45}}{-S_{23}+S_{45}+S_{51}},l_{12}\to \frac{\sqrt{\Delta _1}+S_{12} S_{23}-S_{23} S_{34}+S_{34} S_{45}-S_{12} S_{51}-S_{45} S_{51}}{2 S_{12} S_{23}+2 S_{12} S_{34}-2 S_{12} S_{51}},\nonumber \\
    &l_{13}\to \frac{\sqrt{\Delta _1}+S_{12} S_{23}-S_{23} S_{34}-2 S_{12} S_{45}+S_{34} S_{45}-S_{12} S_{51}-S_{45} S_{51}}{2 S_{12} S_{23}-2 S_{12} S_{45}-2 S_{12} S_{51}},\nonumber \\
    &l_{14}\to \frac{-\sqrt{\Delta _1}-S_{23} S_{34}+S_{34} S_{45}-S_{45} S_{51}-S_{12} \left(S_{51}-S_{23}\right)}{2 S_{12} \left(S_{23}+S_{34}-S_{51}\right)},\nonumber \\
    &l_{15}\to \frac{-\sqrt{\Delta _1}-S_{23} S_{34}+S_{34} S_{45}-S_{45} S_{51}-S_{12} \left(-S_{23}+2 S_{45}+S_{51}\right)}{2 S_{12} \left(S_{23}-S_{45}-S_{51}\right)},l_{16}\to \frac{S_{12} S_{45}-\sqrt{\Delta _2}}{S_{12} S_{34}+S_{12} S_{45}},\nonumber \\
    &l_{17}\to \frac{\sqrt{\Delta _2}+S_{12} S_{45}}{S_{12} S_{34}+S_{12} S_{45}},l_{18}\to \frac{\sqrt{\Delta _3}+S_{12} S_{23}-S_{23} S_{34}-S_{12} S_{45}+S_{34} S_{45}-S_{12} S_{51}-S_{45} S_{51}}{2 S_{12} S_{23}-2 S_{12} S_{45}-2 S_{12} S_{51}},\nonumber \\
    &l_{19}\to \frac{-\sqrt{\Delta _3}-S_{23} S_{34}+S_{34} S_{45}-S_{45} S_{51}-S_{12} \left(-S_{23}+S_{45}+S_{51}\right)}{2 S_{12} \left(S_{23}-S_{45}-S_{51}\right)},l_{20}\to \frac{S_{45}}{S_{12}-S_{34}},\nonumber \\
    &l_{21}\to -\frac{S_{45}}{S_{51}},l_{22}\to \frac{-\sqrt{\Delta _1}-S_{12} S_{23}+S_{23} S_{34}-S_{34} S_{45}+S_{12} S_{51}+S_{45} S_{51}}{2 S_{12} S_{51}},\nonumber \\
    &l_{23}\to \frac{\sqrt{\Delta _1}+S_{23} S_{34}-S_{34} S_{45}+S_{45} S_{51}+S_{12} \left(S_{51}-S_{23}\right)}{2 S_{12} S_{51}},\nonumber \\
    &l_{24}\to \frac{-\sqrt{\Delta _4}+S_{23} S_{34}-S_{34} S_{45}+S_{45} S_{51}+S_{12} \left(-S_{23}+S_{45}+S_{51}\right)}{2 S_{12} S_{51}},\nonumber \\
    &l_{25}\to \frac{\sqrt{\Delta _4}+S_{23} S_{34}-S_{34} S_{45}+S_{45} S_{51}+S_{12} \left(-S_{23}+S_{45}+S_{51}\right)}{2 S_{12} S_{51}},\nonumber \\
    &l_{26}\to \frac{-\sqrt{\Delta _1}+S_{23} S_{34}-S_{34} S_{45}+S_{45} S_{51}+S_{12} \left(-S_{23}+2 S_{45}+S_{51}\right)}{2 S_{12} \left(S_{12}-S_{34}+S_{51}\right)},\nonumber \\
    &l_{27}\to \frac{\sqrt{\Delta _1}+S_{23} S_{34}-S_{34} S_{45}+S_{45} S_{51}+S_{12} \left(-S_{23}+2 S_{45}+S_{51}\right)}{2 S_{12} \left(S_{12}-S_{34}+S_{51}\right)},\nonumber \\
    &l_{28}\to \frac{\sqrt{\Delta _5}+S_{12} S_{45}}{S_{12} \left(S_{45}-S_{23}\right)},l_{29}\to \frac{\sqrt{\Delta _5}-S_{12} S_{45}}{S_{12} \left(S_{23}-S_{45}\right)},l_{30}\to \frac{\left(S_{23}-S_{45}\right) S_{45}}{S_{12} S_{23}+\left(S_{23}-S_{45}\right) S_{45}},\nonumber \\
    &l_{31}\to \frac{-2 S_{45}^3+2 S_{23} S_{45}^2-S_{34} S_{45}^2-S_{51} S_{45}^2+S_{23} S_{34} S_{45}-S_{12} \left(S_{51}-S_{23}\right) S_{45}-\sqrt{\Delta _6}}{2 \left(S_{12} S_{23} \left(S_{34}+S_{45}\right)+\left(S_{23}-S_{45}\right) S_{45} \left(S_{34}+S_{45}\right)-S_{12} S_{45} S_{51}\right)},\nonumber \\
    &l_{32}\to \frac{-2 S_{45}^3+2 S_{23} S_{45}^2-S_{34} S_{45}^2-S_{51} S_{45}^2+S_{23} S_{34} S_{45}+S_{12} \left(S_{23}-S_{51}\right) S_{45}+\sqrt{\Delta _6}}{2 \left(S_{12} S_{23} \left(S_{34}+S_{45}\right)+\left(S_{23}-S_{45}\right) S_{45} \left(S_{34}+S_{45}\right)-S_{12} S_{45} S_{51}\right)}
    \label{eq:alphabet}
\end{align}
with
\begin{align}
   \Delta _1 =& S_{12}^2 \left(S_{23}-S_{51}\right){}^2+\left(S_{23} S_{34}+S_{45} \left(S_{51}-S_{34}\right)\right){}^2 \nonumber \\
   &+2 S_{12} \left(S_{45} S_{51} S_{23}+S_{34} \left(S_{45}+S_{51}\right) S_{23}-S_{23}^2 S_{34}+S_{45} \left(S_{34}-S_{51}\right) S_{51}\right), \\
   \Delta _2 =& S_{12} S_{34} S_{45} \left(-S_{12}+S_{34}+S_{45}\right),\\
   \Delta _3 =& S_{12}^2 \left(-S_{23}+S_{45}+S_{51}\right){}^2+\left(S_{23} S_{34}+S_{45} \left(S_{51}-S_{34}\right)\right){}^2\nonumber \\
   &-2 S_{12} \left(S_{23}-S_{45}-S_{51}\right) \left(S_{23} S_{34}-S_{45} \left(S_{34}+S_{51}\right)\right),\\
   \Delta _4 =& \left(S_{23}^2-2 \left(S_{45}+S_{51}\right) S_{23}+\left(S_{45}-S_{51}\right){}^2\right) S_{12}^2+\left(S_{23} S_{34}+S_{45} \left(S_{51}-S_{34}\right)\right){}^2\nonumber \\
   &-2 \left(S_{34} S_{23}^2+S_{45} S_{51} S_{23}-S_{34} \left(2 S_{45}+S_{51}\right) S_{23}+S_{45} \left(S_{34}-S_{51}\right) \left(S_{45}-S_{51}\right)\right) S_{12},\\
   \Delta _5 =& S_{12} S_{23} \left(S_{12}+S_{23}-S_{45}\right) S_{45},\\
   \Delta _6 =& S_{45}^2 \Delta_1
\end{align}

Each family is characterised by a subset of the full set of letters. In $P_1$ the following 19 letters appear,
\begin{equation}
    \left\{l_1,l_2,l_3,l_4,l_5,l_6,l_7,l_8,l_9,l_{10},l_{11},l_{12},l_{13},l_{14},l_{15},l_{16},l_{17},l_{18},l_{19}\right\},
\end{equation}
in $P_2$ the following 25 letters appear,
\begin{equation}
    \left\{l_1,l_2,l_3,l_4,l_5,l_6,l_7,l_8,l_9,l_{10},l_{11},l_{12},l_{13},l_{14},l_{15},l_{18},l_{19},l_{20},l_{21},l_{22},l_{23},l_{24},l_{25},l_{26},l_{27}\right\},
\end{equation}
and finally in $P_3$ the following 25 letters appear,
\begin{equation}
    \left\{l_1,l_2,l_3,l_4,l_5,l_6,l_7,l_8,l_{10},l_{11},l_{13},l_{15},l_{20},l_{21},l_{22},l_{23},l_{24},l_{25},l_{26},l_{27},l_{28},l_{29},l_{30},l_{31},l_{32}\right\}.
\end{equation}

\bibliographystyle{JHEP}
\bibliography{biblio.bib}

\end{document}